\documentclass[online]{aa-mod}
\usepackage[varg]{txfonts}
\usepackage{natbib}
\usepackage{graphicx}
\usepackage{amsmath}
\usepackage{xcolor}
\usepackage[normalem]{ulem}
\bibpunct{(}{)}{;}{a}{}{,} 

\begin{document} 

\title{A search for millimeter emission from the\\ 
coldest and closest brown dwarf with ALMA}

\authorrunning{D. Petry \& V. D. Ivanov}

   \author{Dirk Petry\inst{1}
           \and
           Valentin D. Ivanov\inst{1}
           }

   \institute{\inst{1}European Southern Observatory, Karl-Schwarzschild-Str. 2,
              85748 Garching, Germany
              \email{dpetry@eso.org}
             }

   \date{Received 5 June 2020; accepted 5 August 2020}

 
\abstract
{\object{WISE\,J085510.83-071442.5} (W0855) is a unique object: with T$_{\rm eff}\approx$250\,K it is the coldest
known brown dwarf (BD), located at
only $\approx$2.2\,pc form the Sun. It is extremely faint, which makes
any astronomical observations difficult. However, at least one remotely similar ultra-low-mass object, the M9 dwarf \object{TVLM\,513-46546}, has been shown to be a
steady radio emitter at frequencies up to 95\,GHz with superimposed active states
where strong, pulsed emission is observed.} 
{Our goal is to determine the millimeter radio properties of W0855 with
deep observations around 93\,GHz (3.2 mm) in order to investigate
whether radio astrometry of this object is feasible and to measure
or set an upper limit on its magnetic field.}
{We observed W0855 for 94\,min at 85.1-100.9\,GHz on 24 December 2019 using 44
of the Atacama Large millimeter Array (ALMA) 12\,m antennas. We used the standard ALMA
calibration procedure and created the final image for our analysis
by accommodating the Quasar 3C\,209, the brightest nearby object by far.
Furthermore, we created a light curve with a 30\,s time resolution to search for pulsed emission.}
{Our observations achieve a noise RMS of 7.3\,$\mu$Jy/beam
for steady emission and of 88\,$\mu$Jy for 30\,s pulses
in the aggregated bandwidth (Stokes I). There is no evidence for
steady or pulsed emission from the object at the time of the observation.
We derive 3\,$\sigma$ upper limits of 21.9\,$\mu$Jy
on the steady emission and of 264\,$\mu$Jy on the pulsed emission
of W0855 between 85~GHz and 101~GHz.
}
{Together with  the recent non-detection of W0855 at 4-8\,GHz,
our constraints on the steady and pulsed emission from W0855 confirm that
the object is neither radio-loud nor magnetospherically particularly active.} 

\keywords{stars: evolution -- brown dwarfs -- stars: individual: WISE\,J085510.83-071442.5 -- stars: magnetic field}
\maketitle

\section{Introduction}

\citet{2013ApJ...767L...1L,2014ApJ...786L..18L} discovered three 
new brown dwarfs (BDs) in two systems (one is a binary) in the 
immediate vicinity of the Sun, and there are reports for other BDs that are
not much more distant
\citep[e.g.][]{2014A&A...561A.113S,2018ApJ...868...44F,2018ApJS..236...28T,2019ApJS..240...19K,2019ApJ...881...17M,2020ApJ...889...74M}. This
indicates that these objects may be very common in the Milky Way. 
As \citet{2014AJ....148...82W} suggested, the all-sky survey 
carried out by the Wide-field Infrared Survey Explorer 
\citep[WISE;][]{2010AJ....140.1868W} may contain more of them. 

The low temperatures of BDs  make them intrinsically extremely faint, even in the mid-infrared (IR), where their spectral 
energy distributions (SEDs) peak. \object{WISE J085510.83$-$071442.5} 
(W0855 hereafter) with T$_\mathrm{eff}$$\approx$250\,K 
\citep{2014ApJ...786L..18L,2014A&A...570L...8B,2019ApJS..240...19K}
is the coldest known BD. Its temperature places 
it at the intersection between giant planets in the Solar System 
(Jupiter is not very different with T$\sim$140-170\,K), cool 
exoplanets at large orbital distances in other planetary systems, 
and the ultracool BDs and free-floating planets at T$\sim$400\,K 
or higher \citep{2013ApJ...764..101B}.

Located as close as $\approx$2.2\,pc to us, W0855 presents a unique
opportunity to study an ultracool object and its environment
in detail. Considerable effort was made to observe it from the 
ground \citep{2014ApJ...797....3K,2014A&A...570L...8B}, but it 
yielded only one marginal detection \citep{2014ApJ...793L..16F}
and the construction of an SED had to wait until some space-based 
observations were collected 
\citep{2013ApJ...767L...1L,2014ApJ...786L..18L,2016AJ....152...78L,2016A&A...592A..80Z}.
They confirmed the initial temperature estimate of 250\,K. 

\citet{2016ApJ...826L..17S} and \citet{2018ApJ...858...97M} 
obtained spectra of W0855 at $\lambda$$\approx$3-4\,$\mu$m. A 
comparison with Jupiter suggests a lack of PH$_3$ absorption 
\citep{1982ApJ...263..443K}. This molecule is unstable; phosphorus is easily captured in P$_4$O$_6$ . The existence 
of PH$_3$ in the Jupiter atmosphere is evidence for vertical 
mixing between the hot interior and cooler outer parts of the planet.
The lack of PH$_3$ in W0855 may indicate that it has a more quiescent 
atmosphere than Jupiter. The reported modelling of these spectra 
suggests that CH$_4$ is present, but at sub-solar abundance.

Low-mass stars and BDs offer a shortcut towards finding small 
planets, comparable to the Earth, because of their advantageous 
planet-to-host mass and luminosity ratios. Not surprisingly, 
the first directly imaged exoplanet orbits a BD 
\citep{2004A&A...425L..29C,2005A&A...438L..25C} and the nearest 
planet was discovered with radial velocity monitoring of the 
M-type star \citep{2016Natur.536..437A}. Hubble Space Telescope ({\it HST)} imaging found 
no companions of W0855 within 0.5\,AU 
\citep{2016AJ....152...78L,2016A&A...592A..80Z} and {\it Spitzer} 
imaging reported none within 9-970\,AU \citep{2015AJ....150...62M}. 
Unfortunately, the object is too faint for radial velocity 
monitoring even with the best current or near-future spectrographs.
This makes astrometry the most promising tool to search for 
companions, likely of planetary mass, around W0855. Because these companions 
do not need to be detected directly, the astrometric method can probe smaller separations than 
direct imaging. \citet{2016AJ....152...78L} monitored W0855 
with {\it Spitzer} \citep[see also ][]{2016AJ....151....9E}.
They measured accurate parallax and proper motion, but found no 
indications for companions.

Another possibility is radio astrometry: Stars and planets are 
well known to emit in the radio. Typically, radio astrometry 
has been applied to evolved or cool and active objects 
\citep[e.g.][]{2018MNRAS.475.1399G,2019ApJ...884...13C}.
\citet{2001Natur.410..338B} opened the field of BD radio 
observations by reporting a radio detection of \object{LP\,944--20} 
\citep[see also][]{2019ApJ...874..136S}. Radio interferometers 
deliver higher angular resolution and 
higher positional accuracy than the optical and IR instruments 
\citep{2019ApJ...875..114X}. These advantages were demonstrated 
in radio-astrometric campaigns on M dwarfs by \citet{2009ApJ...701.1922B},
and on BDs by \citet{2009ApJ...706L.205F}, \citet{2013ApJ...777...70F},
and \citet{2017MNRAS.466.4211G}.

However, radio astrometry is not always feasible. Depending on 
the level of activity, the radio emission can vary widely. This prompted 
us in 2018 to investigate whether W0855 might be accessible by the most sensitive millimeter (mm) 
observatory available today, ALMA\footnote{http://almascience.org}, within a 
reasonable amount of observing time.

The  exact mechanisms of radio emission in ultracool BD are unclear, 
but the best available models \citep[e.g.][]{2012ApJ...760...59N} 
involve strong magnetic fields and plasma outflows. A detection 
would indicate that these can be present in the atmospheres of 
objects in the T$_\mathrm{eff}$$\sim$250\,K regime.

Assuming W0855 is a fast rotator with a strong magnetic field, we 
can estimate its radio properties by scaling the apparent flux of 
potentially similar objects. The only at least remotely similar object with a detection at mm wavelengths is the M9 dwarf 
\object{TVLM\,513-46546} (distance D=10.8\,pc, TVLM\,513 hereafter), that has been detected with ALMA 
at the 56$\pm$12\,$\mu$Jy level by \citet{2015ApJ...815...64W}. It 
would yield an apparent flux of $\approx$1.3\,mJy at 93\,GHz (Band\,3) 
at the distance of W0855. It is considerably hotter than W0855, but the origin of the mm emission is not thermal, as discussed below.
The other example mentioned above, the BD LP944$-$20 \citep[D=6.4\,pc;][]{2001Natur.410..338B}, with an 
apparent flux density of $\approx$80\,$\mu$Jy at 8.5\,GHz, would yield a flux density of $\approx$0.7\,mJy at the distance of W0855. 
These are both optimistic estimates because only a 
fraction of BDs show such a high level of activity. Those that do are 
mostly young, as suggested by the presence of Li in LP944-20 and 
its dusty clouds \citep{2007MNRAS.380.1285P}, but the age of W0855
is still unconstrained.
Fluxes of the order of 1\,mJy are easily within reach: even tentatively
adopting a lower apparent flux by about a factor of 50 (equivalent to a more  
modest activity level) yields a 3\,$\sigma$ detection (i.e. 1\,$\sigma$ 
sensitivity of 0.0087\,mJy) in $\approx$2\,h integration in ALMA 
Band\,3.

A radio-quiet W0855 is beyond reach: an unresolved (the apparent 
angular diameter of a Jupiter-sized BD at the distance of W0855 is 
0.21\,mas) black body of 250\,K at 2.2\,pc has a flux density of 
0.26\,$\mu$Jy at 100\,GHz, requiring a $\approx$2\,yr integration to 
achieve a 3\,$\sigma$ detection.

After the second iteration of our proposal to observe W0855 with 
ALMA had already been approved, however,
 \citet{2019MNRAS.487.1994K} reported for W0855 a 
VLA 4-8\,GHz upper limit on the steady flux of 7.2\,$\mu$Jy.
With this new information, it was unlikely that we would detect steady emission from
W0855 at 93 GHz if its spectrum is similar to that of TVLM\,513, 
which shows a slow decay between 8\,GHz and ALMA Band 3.
However, both TVLM\,513 and other ultracool dwarfs 
have also been reported to have states of high activity where they produce bright broadband pulses of polarised
radio emission \citep[e.g.][]{2007ApJ...663L..25H,2008ApJ...684..644H}
with pulse durations of the order of minutes to hours and substructures down
to the 30\,s level. These are thought to be related to the electron cyclotron maser instability (ECMI)
in the magnetosphere and thus coupled to the rotational periods of the objects. 
The pulses are visible both in full intensity (Stokes~I) and circularly 
polarised emission (Stokes~V) and reach fluxes of 4\,mJy at 8.4\,GHz.
Scaled to the (shorter) distance of W0855 and extrapolated to 93\,GHz 
using the measured spectrum of TVLM\,513, such pulses 
could reach more than 10\,mJy for W0855.
With the sensitivity of ca. 0.1\,mJy of our ALMA observation for 30\,s integration time,
detecting W0855 in such an active state would still be possible.
However, a detection of pulses at 93\,GHz would imply the presence of exceptionally strong magnetic fields in excess of 34\,kG if they were to be explained by the ECMI mechanism.

\section{Observations}

\subsection{Band 3 observations in December 2019}

The observations were carried out between 2019-12-24, 5:01:06~UT
and 2019-12-24, 7:15:23~UT with 44 of the ALMA 12\,m antennas. Two 
ALMA execution blocks with 47\,min on-source time each were carried 
out, yielding a total on-source time of 94\,min. The configuration
was nominally C43-2. The longest baseline was 313\,m, and the 
shortest baseline was 15\,m long. This resulted in a synthesised beam of 
ca. 3$\arcsec$ diameter, which represents our angular resolution (see 
the next section for the exact value). The precipitable water vapour 
(PWV) in the atmosphere above ALMA was between 3.7\,mm and 4.2\,mm 
during the observations (reasonably good values for Band~3).

The observations were scheduled as standard low-resolution (Time 
Division Mode) observations with four spectral windows of 128 channels 
of 15.625\,MHz width each. The first and the last eight~channels of each 
spectral window were discarded for low sensitivity, leaving a
total effective bandwidth of 7\,GHz in the range from 85.1382\,GHz to 
100.8618\,GHz.

We had added 3C\,209 \citep[PKS\,J0855$-$0715, $\alpha$=08:55:09.5, 
$\delta$= $-$07:15:03, J2000;][]{1996AJ....111.1945D} as a check 
source to the observing schedule because the proper motion of W0855 takes it close to this 
bright blazar during ALMA Cycle\,7. Check sources are briefly 
visited during observations for phase and flux calibration. We 
explicitly included the object in the image deconvolution process 
to improve our sensitivity (see next section).

During the observations of the target, the phase centre of the 
interferometer was kept at the position of W0855, taking
the proper motion as measured by \citet{2016AJ....152...78L} into account:
$\alpha$=08:55:05.639915, $\delta$=$-$07:14:36.19233 (J2000) at the 
beginning of the first block execution and $\alpha$=08:55:05.639827, 
$\delta$=$-$07:14:36.19227 (J2000) at the beginning of the second. 
Over the course of the observation, there was therefore no significant 
movement of the target compared to our angular resolution. The target 
elevation was between 56.1$^\circ$ and 74.5$^\circ$.

The check source 3C\,209 was observed during ten 1\,min scans over 
the course of the execution block. Phase, bandpass, and flux calibrators were 
observed following the standard ALMA calibration procedure.
As bandpass and flux calibrator served QSO J0725-0054, 
as phase calibrator QSO J0847-0703. The typical systematic uncertainty on 
the standard ALMA flux calibration is better than 10\% in Band 3 \citep{almatech}.

\subsection{Data analysis}

The calibration of the data followed the standard ALMA Quality 
Assurance procedure for Cycle\,7  \citep[see][]{2014SPIE.9152E..0JP,almatech}
using the calibration pipeline 42866 \citep{almapipe} based on the CASA data 
analysis package version 5.6.1-8 \citep{2019arXiv191209437E}.
The calibrated data were then imaged with the tclean task of 
the same CASA package in mfs mode, that is, combining all spectral 
channels into one image. 
At first, only the W0855 data were imaged as a single field with a 
pixel size of 0.4$\arcsec$ and natural weighting in order to 
optimise the point-source sensitivity. This resulted in a noise 
RMS in the central region of the image of 11.8\,$\mu$Jy, significantly 
worse than the expected 8\,$\mu$Jy.

\begin{figure}[!t]
\centering
\includegraphics[width=9cm]{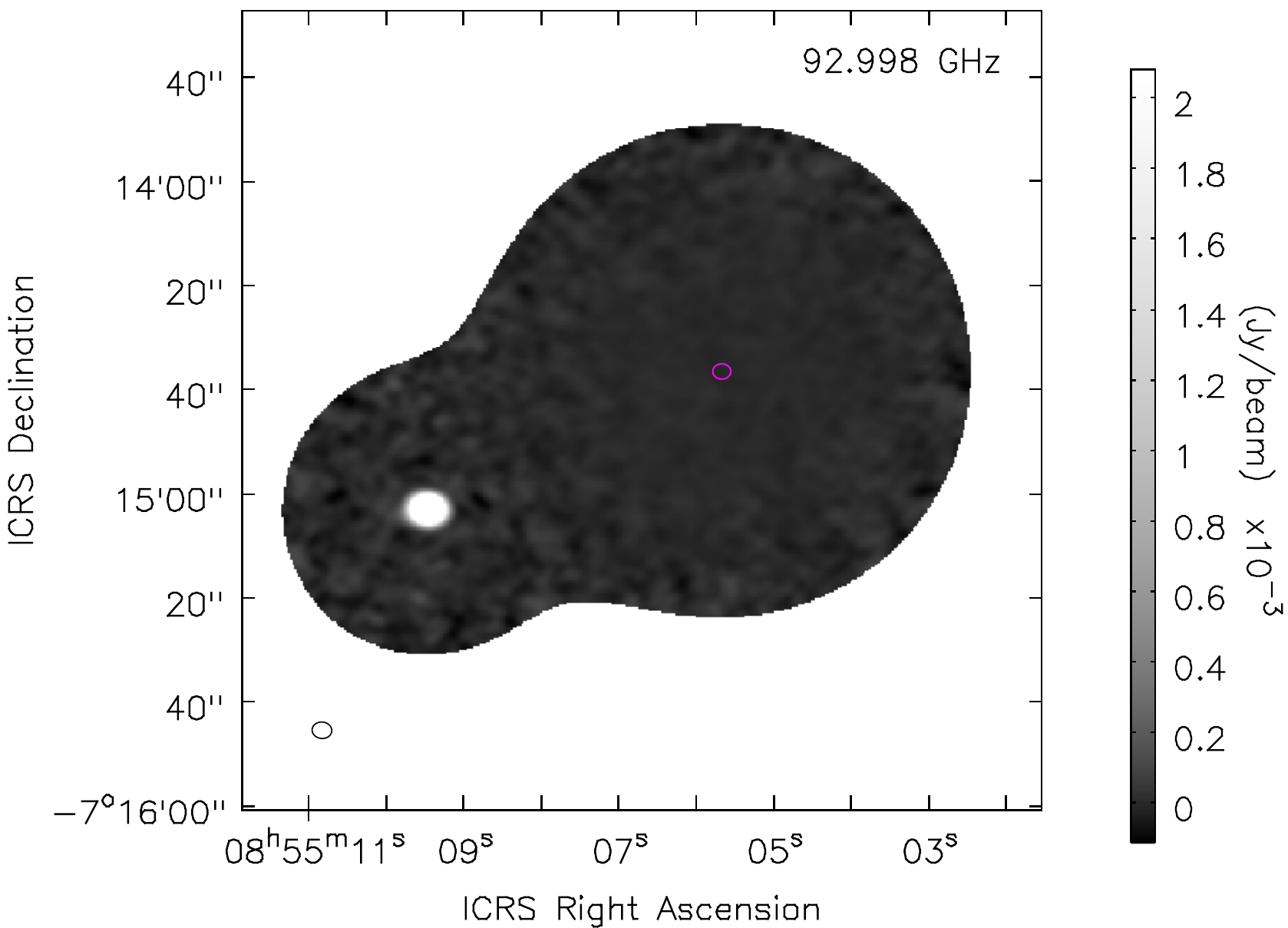}\\
\caption{Map of the W0855 field obtained in this analysis. 
The expected location of the source is indicated with a magenta 
ellipse of the shape of the synthesised beam. The bright source 
to the east is the background quasar 3C\,209 with a flux of ca. 
20\,mJy, i.e. a factor of 2740 brighter than the sensitivity of 
7.3\,$\mu$Jy that is achieved in the vicinity of W0855. To give 
an impression of the homogeneity of the noise, the brightness 
scaling in this high-dynamic-range image was set to strongly 
overexpose the quasar, which therefore appears more extended than 
the beam. The beam size is shown again as an ellipse in the lower 
left corner.}\label{fig:result}
\end{figure}

We suspected that the loss in sensitivity was due to the sidelobes 
of the relatively bright nearby object 3C\,209 and therefore re-imaged the target by combining the check-source field containing 3C\,209
with the target field as a mosaic (but otherwise identical 
imaging parameters). This permitted us to remove the sidelobes of 
3C\,209 in the deconvolution and reduced the noise RMS in the 
region around the nominal position of W0855 to 7.3\,$\mu$Jy. The 
synthesised beam of the observation was measured to be 
3.081$\arcsec$$\times$2.559$\arcsec$. The resulting image is shown 
in Fig.\,\ref{fig:result}.

\subsection{Derivation of an upper limit on the steady flux}

To derive a 3\,$\sigma$ upper limit on the steady (quiescent) 
flux of W0855 between 85.1\,GHz and 100.9\,GHz, 
we follow the prescription of \cite{2019MNRAS.487.1994K} 
(who simply multiplied the measured noise RMS by a factor 3) 
because we wish to plot our result together with theirs;
see \cite{masciul}
for an alternative, more conservative method.
With our noise RMS of 7.3\,$\mu$Jy from above, we 
obtain an upper limit of 21.9\,$\mu$Jy.

\subsection{Search for pulsed emission}

\begin{figure}[!t]
\centering
\includegraphics[width=9cm]{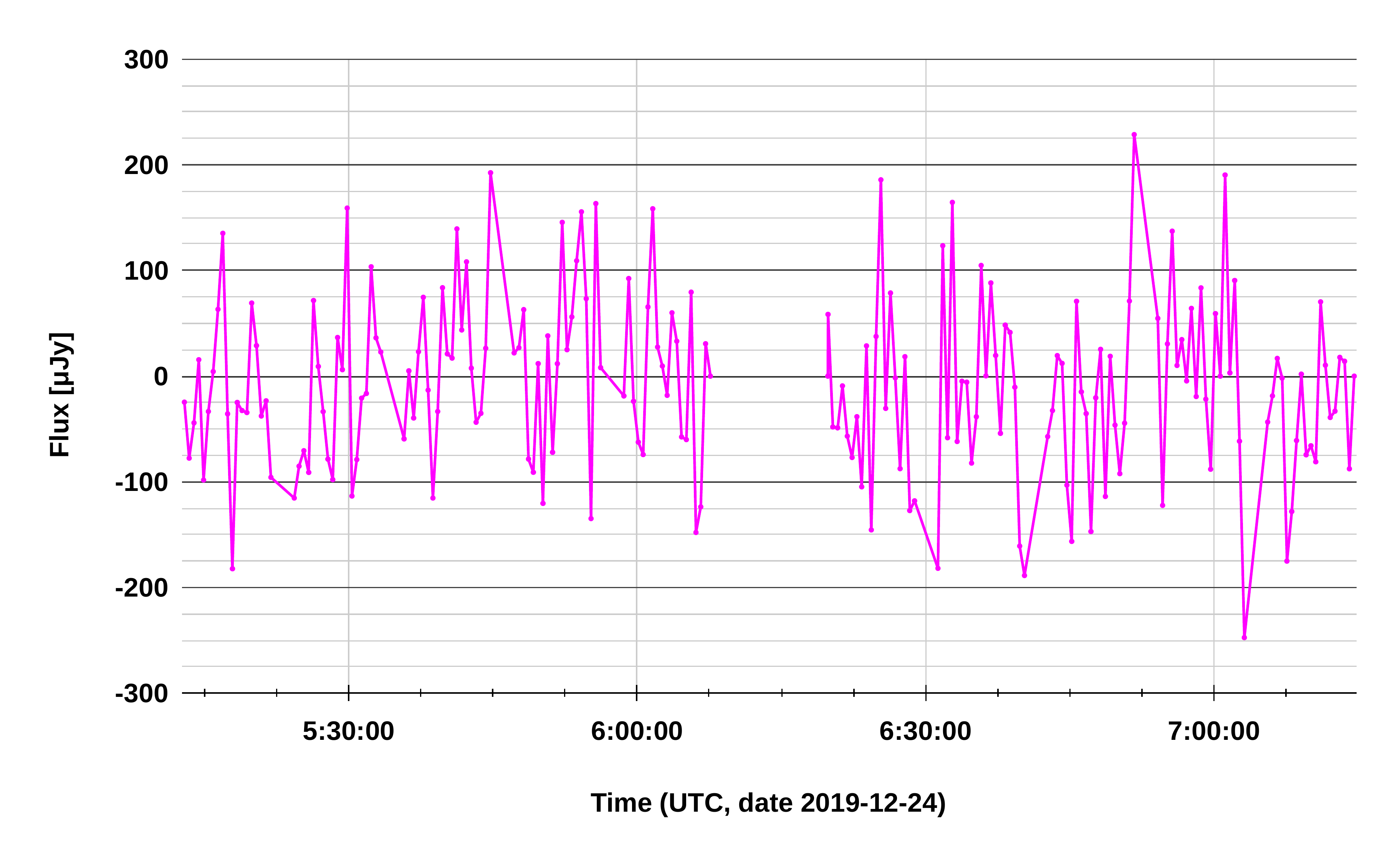}\\
\caption{Light curve of W0855 at 85.1-100.9\,GHz (Stokes I) with 30\,s time resolution obtained in this work. There is no evidence for pulsed emission. The pulse distribution is consistent with a Gaussian and has a mean of -8.6\,$\mu$Jy and an RMS of 82\,$\mu$Jy.}\label{fig:lightcurve}
\end{figure}

Based on the minimum duration of the pulse structures seen in 
TVLM\,513  \citep{2007ApJ...663L..25H}, we divided our
dataset into 30\,s time bins for which we obtained individual
images. Measuring the flux in a beam-shaped region around the
W0855 position in each of the 188~images, we derived a light curve  
(Fig.\,\ref{fig:lightcurve}). The distribution of the 30\,s fluxes is
consistent with a Gaussian with zero average and RMS 82\,$\mu$Jy,
which in turn is consistent with the noise RMS of 88\,$\mu$Jy that
we measured on average in the individual 30\,s images.
A ~3\,$\sigma$ upper limit on the pulsed flux from W0855 between
85.1\,GHz and 100.9\,GHz during the time of the observation can be 
placed at 264\,$\mu$Jy.

\section{Discussion}

\subsection{Radio emission mechanisms in ultracool objects}

The BD LP944-20 \citep[M9, D=6.4\,pc;][]{2019ApJ...874..136S} 
emits at 80\,$\mu$Jy in quiescence and at 3\,mJy at peak flare 
at 8.5\,GHz (where the SED of the flaring emission peaks). This 
is three orders of magnitude too high for stellar coronal 
emission as described by the empirical relation of 
\citet{1993ApJ...405L..63G}, indicating that an additional 
mechanism that powers the radio emission is at work in ultracool 
objects. \citet{2016MNRAS.463.2202B} provided a recent summary: 
the activity indicators in the X-ray and optical suggest a rapid 
activity dropoff at ultracool temperatures, but the radio 
emission is not reduced as much as the emission in other 
wavelength ranges. Radio emission has been detected in 5-10\% of 
the M9-T6 type BDs, which shows a break in the X-ray-to-radio 
relation in the hotter stellar regime. 

The radio detections fall into two categories: 
(1) pulsed, probably rotationally modulated, 100\% polarised, 
probably originating in ECMIs
\citep[][]{2006A&ARv..13..229T}, and 
(2) quiescent non-polarised, always present when pulsed emission 
is detected; a probable origin is depolarised pulsed emission 
\citep{2008ApJ...684..644H} or gyrosynchrotron emission 
\citep{2002ApJ...572..503B}.

Theoretical models by \citet{2012ApJ...760...59N} suggest that 
the ECMI in ultracool objects may originate from an upward
magnetic field component of the magnetosphere-ionosphere coupling, 
flowing as a result of a meridional angular 
velocity gradient in the ionospheric plasma. This is similar 
to the auroral oval in Jupiter \citep{2003JGRA..108.1389G} 
where the flow is driven by the outward diffusion of plasma 
generated by the motion of Io within the magnetosphere of Jupiter 
\citep{1979JGR....84.6554H}.

\subsection{The magnetic field of W0855}\label{sec:SEDs} 

A detection of ECMI-generated radio emission would permit us to set a 
direct {\it \textup{lower}} limit on the magnetic field from the cut-off at the electron cyclotron resonance frequency 
\citep[Eqn.\,1 in][]{2018haex.bookE.171W}.
A VLA {\it \textup{detection}} at 4-8\,GHz implies a magnetic field 
of at least B$\geq$2\,kG \citep{2016ApJ...818...24K}, and 
only LOFAR and GMRT, which work at MHz frequencies, have the potential to probe weaker magnetic fields, but such observations of ultracool BDs have proven to be challenging
\citep{2016MNRAS.463.2202B,2019MNRAS.483..614Z}.
An ALMA detection
at 85-101\,GHz would place a lower limit at B$\geq$34\.kG, an essentially unphysical value for low-mass stars \citep{2007ApJ...656.1121R}. A non-detection of ECMI-like pulses therefore supports the ECMI origin of such pulses seen at lower frequencies in other objects.
For the BD TVLM\,513, a surface magnetic field strength of B$\sim$3\,kG was measured based on its periodic ECMI flares at 4-8\,GHz \citep{2006ApJ...653..690H}. ECMI, however, was ruled out as the origin of the mm emission detected later \citep{2015ApJ...815...64W}. Instead, Williams and collaborators argued that the non-flaring emission at 95\,GHz stems from gyrosynchrotron processes in an ambient magnetic field $\geq$40\,G.

Other direct methods for measuring the magnetic fields of W0855 
are not feasible: we cannot follow 
\citet{2006ApJ...644..497R,2007ApJ...656.1121R} and 
\citet{2017NatAs...1E.184S} to measure the field from line 
broadening because the object is inaccessible in the optical.

\begin{figure}[!t]
\centering
\includegraphics[width=9.0cm]{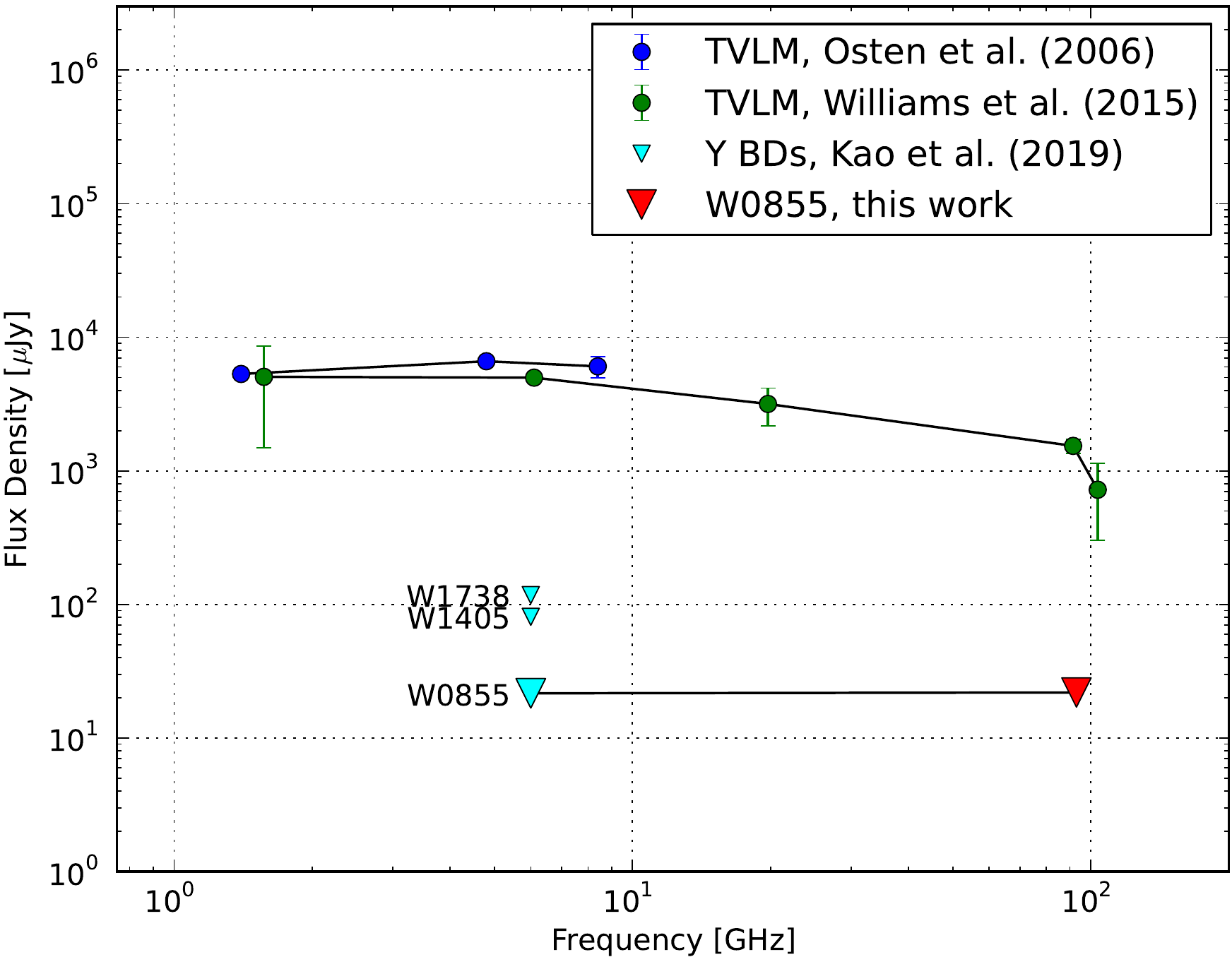}\\
\caption{Radio SED (steady, quiescent emission, Stokes I) of W0855, TVLM\,513-46546 (TVLM), and other 
ultracool dwarfs scaled to the distance of W0855 (D=2.23\,pc). 
See Sec.\,\ref{sec:SEDs} for details.}\label{fig:sed}
\end{figure}

Fig.\,\ref{fig:sed} compares the upper limit on the steady emission from W0855 from 
our work and from \citet{2019MNRAS.487.1994K} with the radio 
SED of the M9 dwarf TVLM\,513
\citep{2006ApJ...637..518O,2015ApJ...815...64W} 
and upper limits for other Y type BDs 
\citep{2019MNRAS.487.1994K}, {\it \textup{all scaled to the distance of 
W0855}}. Clearly, W0855 has the most stringent limit for an 
object of this type.

\section{Summary and conclusions}

We report deep (94 min exposure) ALMA 3~mm observations of 
W0855, an ultracool brown dwarf at $\approx$2\,pc. The object was neither detected as a steady nor as a pulsed emitter. We derive a 3~$\sigma$ upper limit of 21.9\,$\mu$Jy on the steady flux at 85.1-100.9~GHz. 
Likewise, a search for pulsed emission in bins of 30\,s time resolution across our two nearly consecutive 47\,min on-source observation time windows did not result in a detection. We place a 3\,$\sigma$ upper limit on the pulsed Stokes I emission at 264\,$\mu$Jy. Even though both our upper limits are nominally at the same flux level as the corresponding limits derived from the recent non-detection of steady and pulsed emission at 4-8\,GHz by \citet{2019MNRAS.487.1994K},
they are less constraining because the power-law spectrum measured in similar objects such as TVLM\,513 suggests that an index $<$\,0 (see Fig. \ref{fig:sed}) between 8\,GHz and 85\,GHz should be assumed for W0855. 

W0855 is obviously not a prime candidate for the study of magnetospherically active ultracold BDs. Still, because it is so close to Earth, it may be justified to carry out an extremely deep observing campaign with sensitivities at the 0.5\,$\mu$Jy level in order to characterise our unique neighbour. This should first be done in the 4-8\,GHz regime or possibly with the new ALMA Band 1 (35-50\,GHz), which is presently under construction. In addition, longer-term monitoring of this and other ultracool dwarfs at moderate sensitivity is needed to explore variability on timescales longer than a few hours. 

\begin{acknowledgements}
This paper makes use of the following ALMA data: 
ADS/JAO.ALMA\#2019.1.00202.S. ALMA is a partnership of ESO 
(representing its member states), NSF (USA) and NINS (Japan), 
together with NRC (Canada), MOST and ASIAA (Taiwan), and KASI 
(Republic of Korea), in cooperation with the Republic of Chile. 
The Joint ALMA Observatory is operated by ESO, AUI/NRAO and 
NAOJ.
\end{acknowledgements}

\bibliographystyle{aa}

\end{document}